\documentclass[11pt]{article}

\usepackage{graphicx}
\usepackage{amsmath}
\usepackage{amssymb}
\usepackage{amsxtra}

\newcommand{\be}{\begin{equation}}
\newcommand{\en}{\end{equation}}
\newcommand{\bea}{\begin{eqnarray}}
\newcommand{\ena}{\end{eqnarray}}

\title{{\bf How compact stars challenge our view about dark matter}}

\author{Grigoris Panotopoulos \\ grigorios.panotopoulos@tecnico.uçlisboa.pt \\Centro de Astrof{\'i}sica e Gravita\c c\~ao - CENTRA\\Departamento de F{\'i}sica, Instituto Superior T{\'e}cnico-IST,\\ Universidade de Lisboa-UL, Av. Rovisco Pais 1, 1049-001 Lisboa, Portugal}

\begin{document}

\maketitle

\begin{abstract}

It is by now well established that non-relativistic matter in the Universe 
is dominated by dark matter, the origin and nature of which still remains a 
mystery. Although the collisionless dark matter paradigm works very well at large 
distances, a few puzzles at galactic scales arise. These problems may be tackled 
assuming a self-interacting dark matter. If dark matter is accumulated inside a star it will modify its evolution and its properties, such as mass-to-radius profiles and frequency oscillation modes. Asteroseismology is a relatively new, powerful tool that allows us to constrain dark matter models, offering us complementary bounds to the results coming from other means, such as collider or direct searches.
I will present here the main results we have obtained assuming that the
dark matter particle is a boson, which inside a star is modelled as a Bose-Einstein condensate with a polytropic equation-of-state. We have computed i) the radial and non-radial oscillation modes of light clumps of dark matter made of ultra light repulsive scalar fields, and ii) the mass-to-radius profiles as well the frequencies of radial modes of admixed dark matter strange quark stars.
\end{abstract}





\section{Introduction}\label{s:intro}

Since the pioneer work of F.~Zwicky about the dynamics of the Coma galaxy cluster in the 30's \cite{zwicky}, and the observations made by V.~Rubin to determine the rotation curves of galaxies a few decades later \cite{rubin}, we are convinced that most of the non-relativistic matter in the Universe is dark, usually referred to as cold dark matter. In modern times current well-established data coming from many different sources confirm the existence of dark matter \cite{turner}, although its nature and origin still remains a mystery. For a review on dark matter see \cite{DM1,DM2}, and for recent reviews on dark matter detection searches see \cite{DM3,DM4,DM5}. 

Usually dark matter (DM) in the standard parametrization of the Big-Bang cosmological model is assumed to be made of weakly interacting massive particles, a conjecture which works very well at large (cosmological) scales ($\ge Mpc$), but encounters several problems at smaller (galactic) scales, like the core-cusp problem, the diversity problem, the missing satellites problem and the too-big-to-fail problem \cite{2017arXiv170502358T}. These problems may be tackled in the context of self-interacting dark matter \cite{spergel1,spergel2}, as any cuspy feature will be smoothed out by the dark matter collisions. In addition, if dark matter consists of ultralight scalar particles with a mass $m \leq eV$, and with a small repulsive quartic self-interaction a Bose-Einstein condensate (BEC) may be formed with a long range correlation. This scenario has been proposed as a possible solution to the aforementioned problems at galactic scales \cite{proposal1,proposal2,proposal3}. 

Boson stars are star-like, self-gravitating bosonic configurations, where bosons are exclusively trapped in their own gravitational potentials. Boson stars have been studied in \cite{BS1,BS2,BS3,BS4,Mielke2,BS5,BS6,BS7,BS8}, see also \cite{Harko1,Harko2,chavanis1,chavanis2} for Newtonian self-gravitating Bose-Einstein condensates. The maximum mass of bosons stars in non-interacting systems was found in \cite{BS1,BS2}, while in \cite{BS3,BS4} it was shown that self-interactions can cause significant changes. In \cite{BS5,BS6} the authors constrained the boson star parameter space using data from galaxy and galaxy cluster sizes. 

Unlikely many other forms of matter, compact objects \cite{textbook,review1,review2}, which are formed at the end stages of stellar evolution, are unique probes to study the properties of matter under exceptionally extreme conditions. The matter inside such objects is characterized by ultra-high matter densities for which the usual classical description of stellar plasmas in terms of non-relativistic Newtonian fluids is inadequate. Therefore, such very dense compact objects are relativistic and as such, they are only properly described within the framework of Einstein's General Relativity (GR) \cite{GR}.  

Strange quark stars \cite{SS1,SS2,SS3,SS4,SS5,SS6}, at the moment hypothetical objects, can be viewed as ultra-compact NSs. Since quark matter is by assumption absolutely stable, it may be the true ground state of hadronic matter \cite{witten,farhi}, and therefore this new class of relativistic compact objects has been proposed as an alternative to typical NSs. In fact strange quark stars may explain the observed super-luminous supernovae \cite{SL1,SL2}, which occur in about one out of every 1000 supernovae explosions, and which are more than 100 times more luminous than regular supernovae. One plausible explanation is that since quark stars are much more stable than NSs, they could explain the origin of the huge amount of energy released in super-luminous supernovae. Many works have been recently proposed to validate its existence in different astrophysical scenarios \cite{PL18a,Mukhopadhyay16}. 

It is well-known that the properties of stars, such as mass and radius, depend crucially on the equation-of-state. Furthermore,the presence of DM inside a star is expected to influence the structure, the evolution as well as certain properties of the object, such as mass-to-radius profiles and frequency oscillation modes. Even if dark matter does not interact directly with normal matter, it can have significant gravitational effects on stellar objects  DM that can influence evolution and structure of compact objects \cite{maxim1,maxim2,maxim3,maxim4,fermion1,fermion2,boson3,massiveNS,chinos,admixed,fermion5,ellis1,ellis2,extra,GPindios}. Given the recent advances in Helioseismology and Asteroseismology in general, studying the oscillations of stars and computing the frequency modes offer us the opportunity to probe the interior of the stars and learn more about the equation-of-state, since the precise values of the frequency modes are very sensitive to the thermodynamics of the internal structure of the star \cite{2012RAA....12.1107T}. For previous works on radial oscillations of stars see \cite{Cox,Frandsen,Aerts,Hekker,kokkotas,hybrid,basic,LP,fermion3} and references therein.


\section{Impact of DM on strange quark stars}

In the first part of the presentation we discuss the impact of bosonic self-interacting DM on properties of strange quark stars.

\subsection{Mass-to-radius profiles}

-{Structure equations}: We briefly review relativistic stars in General Relativity (GR). The starting point is Einstein's
field equations without a cosmological constant
\be
G_{\mu \nu} = R_{\mu \nu}-\frac{1}{2} R g_{\mu \nu}  = 8 \pi T_{\mu \nu}
\en
where we have set Newton's constant equal to unity, $G=1$, and in the exterior problem the matter energy momentum
tensor vanishes. For matter we assume a perfect fluid with pressure $p$, energy density $\rho$ and an equation of state
$p(\rho)$, while the energy momentum trace is given by $T=-\rho+3p$.
For the metric in the case of static spherically symmetric spacetimes we consider the following ansatz
\be
ds^2 = -f(r) dt^2 + g(r) dr^2 + r^2 d \Omega^2
\en
with two unknown functions of the radial distance $f(r), g(r)$. For the exterior problem one obtains the well-known Schwarzschild solution \cite{SBH}
\be
f(r) = g(r)^{-1} = 1-\frac{2 M}{r}
\en
where $M$ is the mass of the star.
For the interior solution we introduce the function $m(r)$ instead of the function $g(r)$ defined as follows
\be
g(r)^{-1} = 1-\frac{2 m(r)}{r}
\en
so that upon matching the two solutions at the surface of the star we obtain $m(R)=M$, where $R$ is the radius of the star.
The Tolman-Oppenheimer-Volkoff (TOV) equations for the interior solution of a relativistic star with a
vanishing cosmological constant read \cite{OV,tolman}
\bea
m'(r) & = & 4 \pi r^2 \rho(r) \\
p'(r) & = & - (p(r)+\rho(r)) \: \frac{m(r)+4 \pi p(r) r^3}{r^2 (1-\frac{2 m(r)}{r})}
\ena
where the prime denotes differentiation with respect to r, and the equations are to be integrated with the initial conditions
$m(r=0)=0$ and $p(r=0)=p_c$, where $p_c$ is
the central pressure. The radius of the star is determined requiring that the pressure vanishes at the surface,
$p(R) = 0$, and the mass of the star is then given by $M=m(R)$.

-Two-fluid formalism: Now let us assume that the star consists of two fluids, namely strange matter (de-confined quarks) and dark matter with only gravitational interaction between them, and equations of state $p_s(\rho_s)$, $p_\chi(\rho_\chi)$ respectively. The total pressure and the total energy density of the system are given by $p=p_s+p_\chi$ and $\rho=\rho_s+\rho_\chi$ respectively.
Since the energy momentum tensor of each fluid is separately conserved, the TOV equations in the two-fluid formalism for the interior solution
of a relativistic star with a vanishing cosmological constant read \cite{2fluid1,2fluid2}
\bea
m'(r) & = & 4 \pi r^2 \rho(r) \\
p_s'(r) & = & - (p_s(r)+\rho_s(r)) \: \frac{m(r)+4 \pi p(r) r^3}{r^2 (1-\frac{2 m(r)}{r})} \\
p_\chi'(r) & = & - (p_\chi(r)+\rho_\chi(r)) \: \frac{m(r)+4 \pi p(r) r^3}{r^2 (1-\frac{2 m(r)}{r})}
\ena
In this case in order to integrate the TOV equations we need to specify the central values both for normal matter and for
dark matter $p_s(0)$ and $p_\chi(0)$ respectively. So in the following we show the mass-radius diagram for a certain value of the constant $K=2 \pi l/m_\chi^3$ and for fixed dark matter fraction
\be
\epsilon = \frac{p_\chi(0)}{p_s(0)+p_\chi(0)}
\en
and we consider four cases, namely $\epsilon=0.02, 0.035, 0.05, 0.09$. We have chosen these values in agreement with the current dark matter constraints 
obtained from stars like the Sun. Actually, as shown by several authors, even smaller amounts of DM (as a percentage of the total mass of the star) can have a quite visible impact on the structure of these stars~\cite{LopesSilk,ilidio2,ilidio5}. 
As we discuss in this work even such small amounts of DM can change the $M-R$ relation of neutron stars. 

\begin{figure}[ht!]
\centering
{\hspace{-1.0cm}
\includegraphics[scale=0.70]{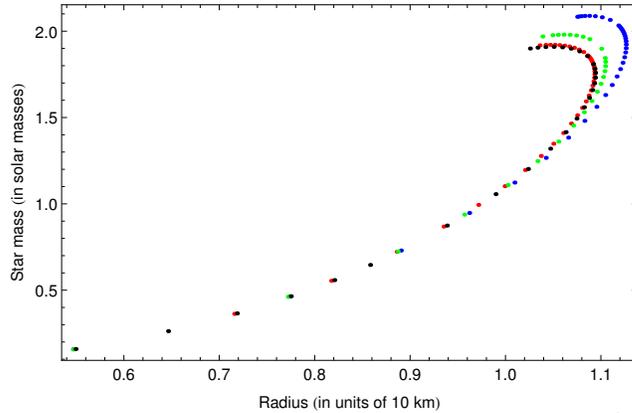} 
\vspace{-0.5cm}
}
\caption{Mass-to-radius profile for $K=4/B$.}
\label{fig:1} 	
\end{figure}

\begin{figure}[ht!]
\centering
{\hspace{-1.0cm}
\includegraphics[scale=0.70]{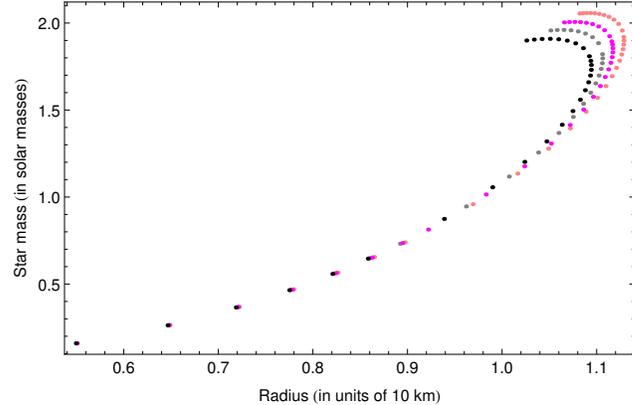}
\vspace{-0.5cm}
}
\caption{Mass-to-radius profile for $K=150/B$.}
\label{fig:2} 	
\end{figure}

-Equation-of-states: For the condensed dark matter we shall consider the equation of state obtained in \cite{darkstars}, namely $P_\chi=K \rho_\chi^2$,
where the constant $K=2 \pi l/m_\chi^3$ is given in terms of the mass of the dark matter particles $m_\chi$ and the scattering length $l$.
In a dilute and cold gas only the binary collisions at low energy are relevant, and these collisions are characterized by the s-wave scattering length
$l$ independently of the form of the two-body potential \cite{darkstars}. Therefore we can consider a short range repulsive delta-potential of the form
\begin{equation}
V(\vec{r}_1-\vec{r}_2) = \frac{4 \pi l}{m_\chi} \delta^{(3)}(\vec{r}_1-\vec{r}_2)
\end{equation}
which implies a dark matter self interaction cross section of the form $\sigma_\chi=4 \pi l^2$ \cite{chinos,darkstars}. Following previous studies
we fix the scattering length to be $l=1 fm$ \cite{chinos,darkstars}, and for $\sigma_\chi/m_\chi$ we apply the bounds discussed in the Introduction
\begin{equation}
0.45 \frac{cm^2}{g} < \frac{\sigma_\chi}{m_\chi} < 1.5 \frac{cm^2}{g}
\end{equation}
which then implies the following range for the mass of the dark matter particle
\begin{equation}
0.05 GeV < m_\chi < 0.16 GeV
\end{equation}
and thus for the constant $K$
\begin{equation}
\frac{4}{B} < K < \frac{150}{B}
\end{equation}
where now the constant $K$ is given in units of the bag constant. Our main results are shown in figures \ref{fig:1} and \ref{fig:2}. In Fig.~\ref{fig:1} we show the mass-to-radius profiles for $K=4/B$ and for $\epsilon=0.02, 0.05, 0.09$, while in Fig.~\ref{fig:2} we show the profiles for $K=150/B$ and for $\epsilon=0.02, 0.035, 0.05$. The standard curve corresponding to no DM (in black) is shown as well for comparison reasons.

\smallskip

For strange matter we shall consider the simplest equation of state corresponding to a relativistic gas of de-confined quarks,
known also as the MIT bag model \cite{bagmodel1,bagmodel2}
\be
p_s = \frac{1}{3} (\rho_s - 4B)
\en
and the bag constant has been taken to be $B=(148 MeV)^4$ \cite{Bvalue}.

\subsection{Radial oscillations}

If $\delta r$ is the radial displacement and $\delta P$ is the perturbation of the pressure, the equations governing the dimensionless
quantities $\xi=\delta r/r$ and $\eta=\delta P/P$ are the following \cite{chanmugan1,chanmugan2}
\be
\xi'(r) = -\frac{1}{r} \left( 3 \xi + \frac{\eta}{\gamma} \right) - \frac{P'}{P+\epsilon} \xi
\en

\be
\begin{split}
\eta'(r) = \xi \left[ \omega^2 r \frac{P+\epsilon}{P}
e^{\lambda-A} - \frac{4 P'}{P} - 8 \pi (P+\epsilon)  r e^{\lambda} + \frac{r (P')^2}{P (P+\epsilon)}\right] \\
&+ \eta \left[ -\frac{\epsilon P'}{P (P+\epsilon)}-4 \pi (P+\epsilon) r e^{\lambda} \right]
\end{split}
\en
where  $e^{\lambda}, e^{A}$ are the two metric functions, $\omega$ is the frequency oscillation mode, and $\gamma$ is the relativistic
adiabatic index defined to be
\begin{equation}
\gamma = \frac{d P}{d \epsilon} (1+\epsilon/P)
\end{equation}

\begin{figure}[ht!]
\centering
{\hspace{-1.0cm}
\includegraphics[scale=0.70]{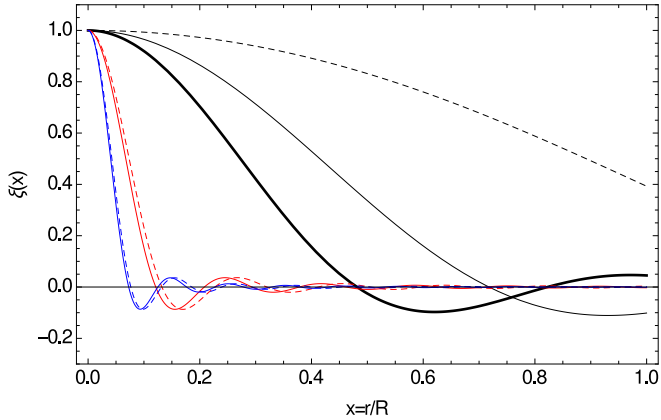} 
\vspace{-0.5cm}
}
\caption{Eigenfunctions $\xi$ vs $r/R$.}
\label{fig:3} 	
\end{figure}

\begin{figure}[ht!]
\centering
{\hspace{-1.0cm}
\includegraphics[scale=0.70]{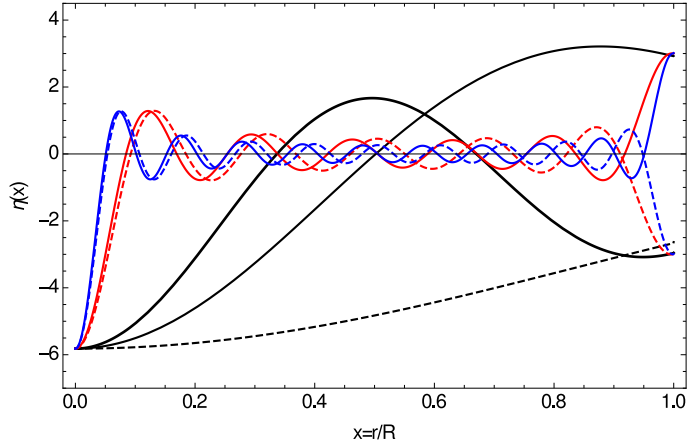} 
\vspace{-0.5cm}
}
\caption{Eigenfunctions $\eta$ vs $r/R$.}
\label{fig:4} 	
\end{figure}

\begin{figure}[ht!]
\centering
{\hspace{-1.0cm}
\includegraphics[scale=0.70]{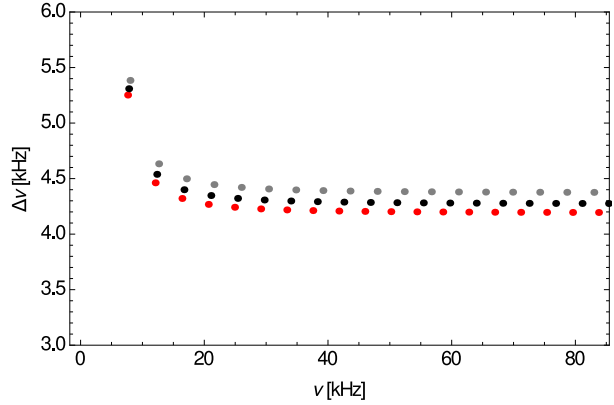} 
\vspace{-0.5cm}
}
\caption{Large frequency separation.}
\label{fig:5} 	
\end{figure}

The system of two coupled first order differential equations is supplemented with two boundary conditions, one at the center as $r \rightarrow 0$, and
another at the surface $r=R$.
The boundary conditions are obtained as follows: In the first equation, $\xi'(r)$ must be finite as $r \rightarrow 0$, and therefore we require
that
\be
\eta = -3 \gamma \xi
\en
must satisfied at the center. Moreover, in the second equation, $\eta'(r)$ must be finite at the surface as $\epsilon,P \rightarrow 0$ and therefore
we demand that
\be
\eta = \xi \left[ -4 + (1-2M/R)^{-1}  \left( -\frac{M}{R}-\frac{\omega^2 R^3}{M} \right )  \right]
\en
must satisfied at the surface, where we recall that $M,R$ are the mass and the radius of the star respectively. Using the shooting method we first compute
the dimensionless quantity $\bar{\omega} = \omega t_0$ where $t_0=1 ms$. Then the frequencies are computed by
\be
\nu = \frac{\bar{\omega}}{2 \pi} \; kHz
\en
Therefore, contrary to the previous hydrostatic equilibrium problem, which is an initial value problem, this is a Sturm-Liouville boundary value problem,
and as such the frequency $\nu$ is allowed to take only particular values, the so-called eigenfrequencies $\nu_n$. Each $\nu_n$ corresponds to a specific radial oscillation mode of the star. Accordingly, each radial mode is identified by its $\nu_n$ and  by an associated pair of eigenfunctions -- the displacement perturbation $\xi_n(r)$  and the pressure perturbation $\eta_n(r)$. Our main results are shown in figures \ref{fig:3}, \ref{fig:4} and \ref{fig:5}. In particular, in Fig.~\ref{fig:3} we show several eigenfunctions $\xi_n$ (n=0,1,2,10,11,18,19) versus normalized coordinate distance $r/R$, in Fig.~\ref{fig:4} we show several eigenfunctions $\eta_n$ (same values of $n$), and in Fig.~\ref{fig:5} we show the large frequency separation $\Delta \nu_n = \nu_{n+1} - \nu_n$ versus frequencies in $kHz$ for 3 cases, namely i) no DM (red color), $5\%$ of DM (black color) and $12\%$ of DM (grey color).

\section{Newtonian stars made of ultralight repulsive DM}

Equation-of-state: The perturbative Lagrangian of a relativistic real scalar field $\phi$ is given by 
\begin{equation}
\mathcal{L} = \frac{1}{2} \partial_\mu \phi \partial^\mu \phi - V(\phi)
\end{equation}
where the scalar potential is of the form
\begin{equation}
V(\phi) = \frac{1}{2} m^2 \phi^2 + \frac{1}{24} \: \frac{m^2}{F^2} \phi^4  + ...
\end{equation}
and where we consider renormalizable theories only, ignoring all higher order terms. In this work the scalar field is identified with any pseudo-Goldstone boson. The sign of the quartic self-interaction is taken to be positive since we assume a repulsive self-interaction for the Dark pseudo-Goldstone boson. Therefore, the model assumed here is characterized by two unknown mass scales, namely the mass of the scalar particle, $m$, as well as the decay constant, $F \gg m$, arising from the spontaneous breaking of some global symmetry. Unfortunately, it turns out that it is not easy to obtain scalar field models with a tiny mass and a repulsive force within known Particle Physics, although some attempts have been made \cite{Fan}. In the following, without relying on concrete Particle Physics models, we shall assume that this is possible, and we shall study radial oscillations of objects made of Dark pseudo-Goldstone bosons.

The above scalar potential combined with the Gross-Pitaevskii equation \cite{BEC1,BEC2,BEC3}, also known as non-linear Schr{\"o}dinger equation, leads to the following equation-of-state for the ultralight 
pseudo-Goldstone boson \cite{Fan,chavanis3}:
\begin{equation}
P(\epsilon) = K \epsilon^2
\end{equation}
where the constant $K$ is computed to be \cite{Fan,chavanis3}
\begin{equation}
K = \frac{1}{(2 \Lambda)^4}
\end{equation}
where a new mass scale $\Lambda \equiv \sqrt{m F}$ has been introduced.

-Hydrostatic equilibrium: Since the axion star is non-relativistic described by a polytropic EoS, to study the hydrostatic equilibrium one has to solve the non-relativistic version of the Tolman-Oppenheimer-Volkoff (TOV) equations \cite{OV,tolman}
\begin{equation}
m'(r) = 4 \pi r^2 \epsilon(r)
\end{equation}
for the mass function, and
\begin{equation}
P'(r) = - \epsilon(r) \frac{m(r)}{r^2}
\end{equation}
for the pressure, where the prime denotes differentiation with respect to the radial coordinate $r$.
Combining these two equations we can derive a single second order differential equation, known as the
Lane-Emden equation \cite{textbook}
\begin{equation}
\frac{d}{dx} \left(x^2 \frac{d \theta}{dx} \right) = -x^2 \theta
\end{equation}
with the initial conditions
$ \theta(0)  =  1  $ and $
{d \theta}/{dx}(0)  =  0$,
where the new variables are defined as follows
\begin{eqnarray}
x & = & \frac{r}{a} 
\end{eqnarray}
and
\begin{eqnarray}
\theta & = & \frac{\epsilon}{\epsilon_c}
\end{eqnarray}
with $\epsilon_c$ being the central energy density, while $a$ is given by
$a = \sqrt{K/2 \pi}$.
It is easy to verify that the solution
\begin{equation}
\theta(x) = \frac{sin(x)}{x}
\end{equation}
satisfies both the Lane-Emden equation and the initial conditions.
Therefore, the energy density as a function of the radial coordinate is given by
\begin{equation}
\epsilon(r) = \epsilon_c \frac{sin(r/a)}{(r/a)}.
\end{equation}
The above equation is valid for the radius varying from $r=0$ until the first zero of the function $\epsilon(r) $, therefore the function $\epsilon(r) $ varies between $\epsilon_c$ and 0. Finally, the mass $M$ and the radius $R$ of the star are given by
\begin{eqnarray}
M & = & 4 \pi \epsilon_c a^3 \int_0^\pi dx x^2 \theta(x) \\
R & = & \pi a
\end{eqnarray}
Clearly, only the mass of the star depends on the central energy density, while the radius is fixed. This happens only in the special case $n=1$, whereas in general both $M$ and $R$ depend on the central energy density. This can also be seen in the mass-to-radius profile for Newtonian boson stars with repulsive forces as shown 
in the work of Chavanis and collaborators (see Fig.~2 of \cite{chavanis1} and Fig.~4 of \cite{chavanis2}).

\begin{figure}[ht!]
\centering
{\hspace{-1.0cm}
\includegraphics[scale=0.70]{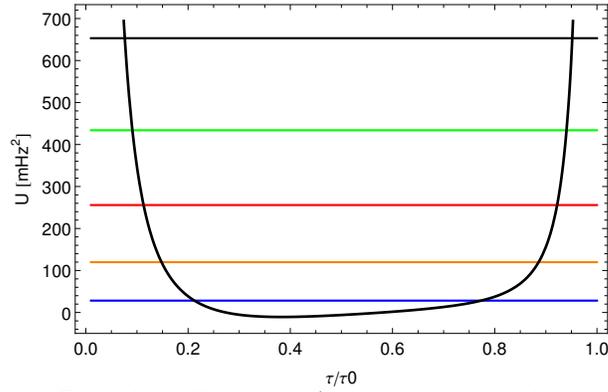} 
\vspace{-0.5cm}
}
\caption{Radial oscillations: Acoustic potential vs acoustic time.}
\label{fig:6} 	
\end{figure}

\begin{figure}[ht!]
\centering
{\hspace{-1.0cm}
\includegraphics[scale=0.70]{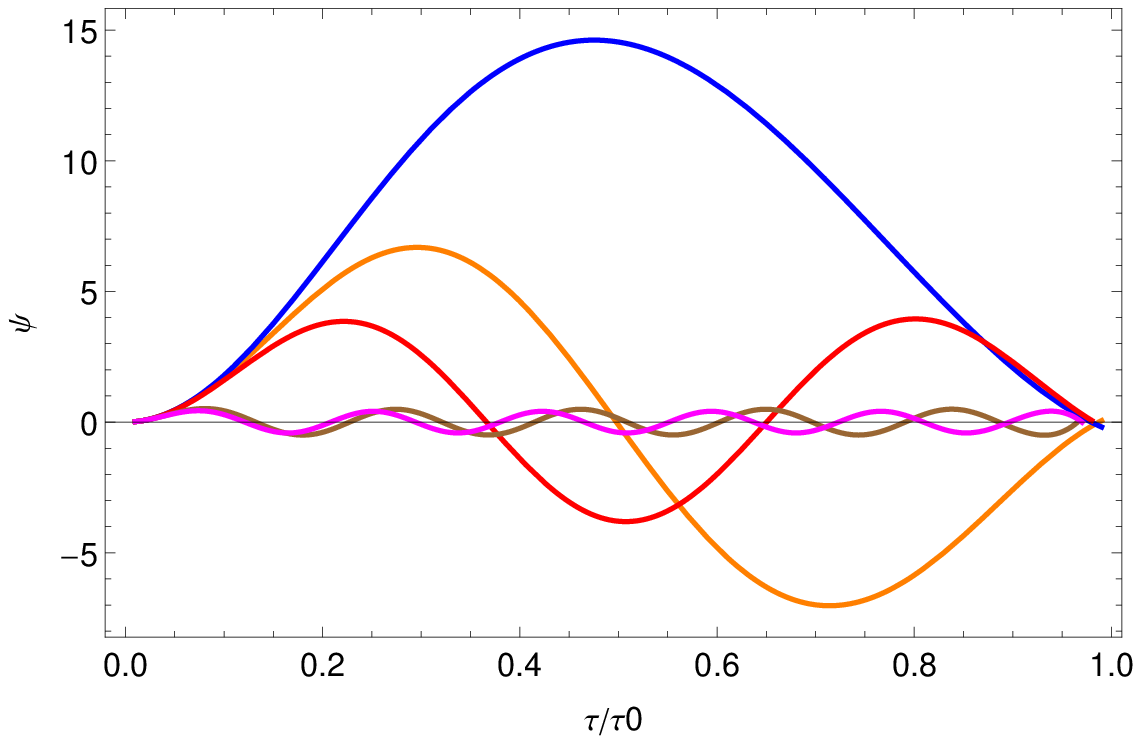} 
\vspace{-0.5cm}
}
\caption{Radial oscillations: Eigenfunctions vs acoustic time.}
\label{fig:7} 	
\end{figure}

\begin{figure}[ht!]
\centering
{\hspace{-1.0cm}
\includegraphics[scale=0.90]{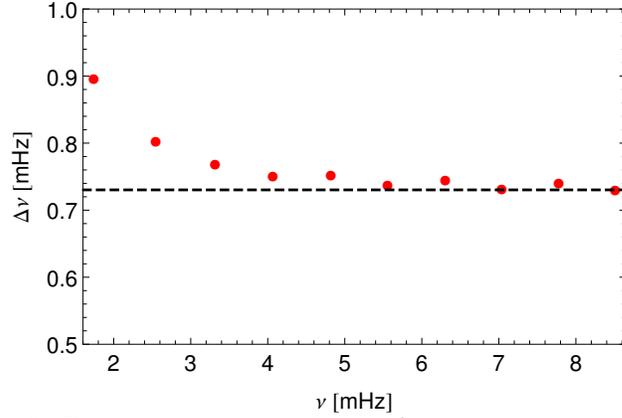} 
\vspace{-0.5cm}
}
\caption{Radial oscillations: Large frequency separation in $mHz$.}
\label{fig:8} 	
\end{figure}

\subsection{Radial oscillations}

In the first part of the presentation we presented the first order system of two coupled equations for the perturbations of a pulsating star. Here, however, we prefer to work equivalently with a second order differential equation used in \cite{basic}
\begin{equation}
-f^2 \zeta'' + G \zeta' + (H-\omega^2) \zeta = 0,
\end{equation}
supplemented with the boundary conditions at the origin $r=0$ and at the surface of the star $r=R$: $\zeta(r=0) = 0$ and $\delta p(r=R) = 0$. In the previous equation $\zeta=r^2 e^{-A/2} \xi$, with $e^{A} \equiv g_{tt}$ being the temporal component of the metric tensor, while the background functions $f$,  $G$ an $H$  are given by
\begin{equation}
f^2(r) =  \frac{\gamma P e^{A-\lambda}}{P+\epsilon}, 
\end{equation}

\begin{equation}
G(r) =   -\frac{f^2}{\gamma P}
\left[ \frac{\gamma P}{2} (\lambda+3A) + (\gamma P)' - \frac{2 \gamma P}{r} \right], 
\end{equation}

\begin{equation}
H(r) = - \frac{f^2}{\gamma P} \left[ \frac{4 P'}{r} + 8 \pi P (P+\epsilon) e^\lambda - \frac{(P')^2}{P+\epsilon} \right], 
\end{equation}
and finally  the  perturbation of the pressure 
can be computed as 
\begin{equation}
\delta p(r) = -\frac{e^{A/2}}{r^2} ( \zeta P' + \gamma P \zeta' ).
\end{equation}

The Sturm-Liouville boundary value problem at hand can be treated equivalently as a quantum mechanical problem by recasting the second order differential equation for $\zeta$ into a Schr{\"o}dinger-like equation \cite{Lopes1} of the form
\begin{equation}
\frac{d^2 \psi}{d \tau^2} + \left[ \omega^2 - U(\tau) \right] \psi = 0,
\end{equation}
where the new variables $\tau$ and $ \psi$ are defined as acoustic radius
$ \tau  =  \int_0^r f^{-1}(z) dz$ and  $ \psi (\tau)  = \zeta/u$.
The effective potential is found to be
\begin{equation}
U = H + \frac{\Pi^2}{4}+\frac{f \Pi'}{2},
\end{equation}
where the function $\Pi$ is given by
$ \Pi  =  -f'-G/f $
while $u$ is determined by the condition
$ u'/u= -\Pi/(2f)$.

The acoustic potential with the first 5 eigenvalues, the corresponding eigenfunctions as well as the large frequency separation in $mHz$ are shown in the figures \ref{fig:6}, \ref{fig:7} and \ref{fig:8}, respectively.

\subsection{Nonradial oscillations}

Linear adiabatic acoustic perturbations in the Cowling approximation \cite{cowling}, where the perturbations of the gravitational potential are neglected, are described by the following equation \cite{ledoux}
\begin{equation}
\zeta''(r) + \left( \frac{2}{r} + \frac{2 \epsilon'(r)}{\epsilon(r)} \right) \zeta'(r) + \left(\frac{\omega_{n,l}^2}{c_s^2} - \frac{l (l+1)}{r^2} \right)  \zeta(r) = 0
\end{equation}
where 
 $c_s$ is the speed of sound defined by $c_s^2 = dP/d\epsilon$, $\omega_{n,l}$ ($=2\pi\nu_{n,l}$) are the discrete eigenvalues, $l > 0$ is the degree of angular momentum (or degree of the mode), and $n=0,1,2,...$ is the overtone number (or radial mode).

The Sturm-Liouville boundary value problem at hand can be treated equivalently as a quantum mechanical problem by recasting the second order differential equation for $\zeta$ into a Schr{\"o}dinger-like equation \cite{Lopes1,Lopes2,Lopes3} of the form
\begin{equation}
\frac{d^2 \psi}{d \tau^2} + \left[ \omega^2 - U_l(\tau) \right] \psi = 0.
\label{eq:Psi}
\end{equation}
Introducing the functions
\begin{equation}
A(r) = \frac{2}{r} + \frac{2 \epsilon'(r)}{\epsilon(r)},
\end{equation}
\begin{equation}
H_l(r) = c_s(r)^2 \frac{l (l+1)}{r^2},
\end{equation}
and
\begin{equation}
P(r) = A(r) c_s(r) - c_s'(r).
\end{equation}
The new variables $\tau$ and $\psi$ are defined as follows
\begin{equation}
\psi(r)  = \frac{\zeta(r)}{u(r)}
\end{equation}
where $u$ satisfies the condition $u'/u= -P/(2 c_s)$, and $\tau$ is the acoustic time
\begin{equation}
\tau  =  \int_0^r c_s^{-1}(z) dz.
\end{equation}
Finally, the effective potential is found to be
\begin{equation}
U_l(r) = H_l(r) + \left( \frac{P(r)}{2} \right)^2+\frac{c_s(r) P'(r)}{2},
\label{eq:Ul}
\end{equation}
and we thus obtain the effective potential as a function of the acoustic time in parametric form $\tau(r), U_l(r)$. 

\begin{figure}[ht!]
\centering
{\hspace{-1.0cm}
\includegraphics[scale=0.80]{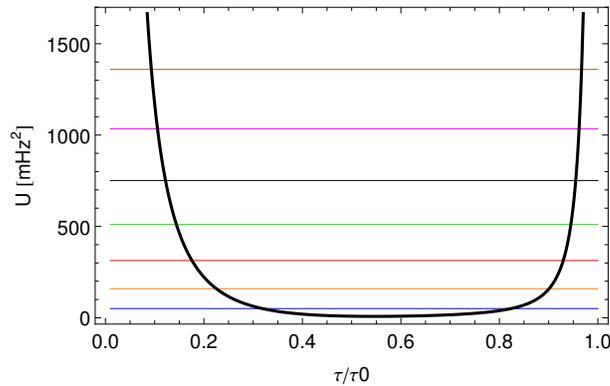} 
\vspace{-0.5cm}
}
\caption{Non radial oscillations: Acoustic potential vs acoustic time for $l=2$.}
\label{fig:9} 	
\end{figure}

\begin{figure}[ht!]
\centering
{\hspace{-1.0cm}
\includegraphics[scale=0.70]{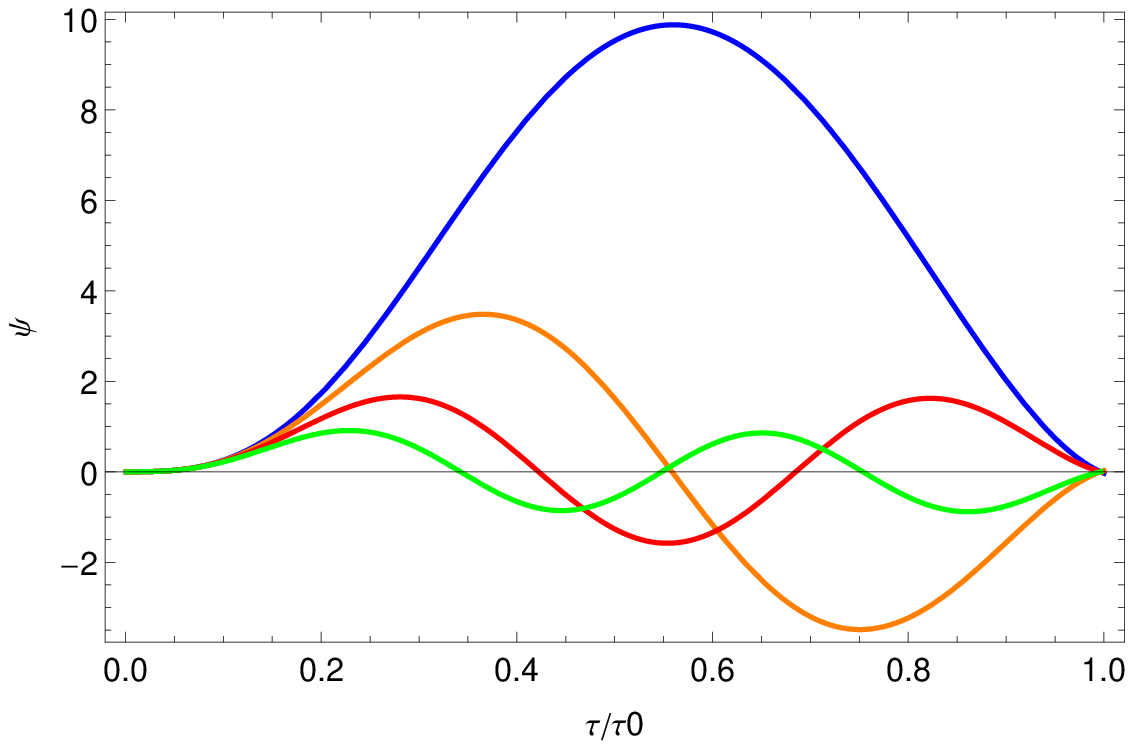} 
\vspace{-0.5cm}
}
\caption{Non radial oscillations: Eigenfunctions vs acoustic time for $l=2$.}
\label{fig:10} 	
\end{figure}

\begin{figure}[ht!]
\centering
{\hspace{-1.0cm}
\includegraphics[scale=0.75]{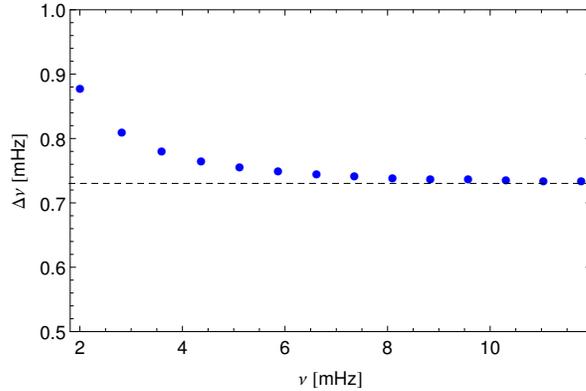} 
\vspace{-0.5cm}
}
\caption{Non radial oscillations: Large frequency separation in $mHz$ for $l=2$.}
\label{fig:11} 	
\end{figure}

The acoustic potential with the first 7 eigenvalues, the corresponding eigenfunctions as well as the large frequency separation in $mHz$ for $l=2$ are shown in the figures \ref{fig:9}, \ref{fig:10} and \ref{fig:11}, respectively.

\section{Conclusions}

In this presentation we have presented results of our work on properties of self-interacting scalar field dark matter in two respects. In particular, in the first part we studied the impact of dark matter on the mass-to-radius profiles as well as on the radial oscillation modes of non-rotating, spherically symmetric strange quark stars in which dark matter is accumulated. Then, in the second part we studied radial and non-radial oscillations of self-gravitating bosonic (star-like) configurations. 

Strange stars are hypothetical compact objects that are supposed to be much more stable than neutron stars, and thus could explain the super luminous supernovae.
For the star interior problem we have solved numerically the Tolman-Oppenheimer-Volkoff equations in the two-fluid formalism. For strange matter we have assumed the simplest version of the MIT bag model (radiation plus the bag constant), while if dark matter is modelled inside the star as a BEC, it can be described by a polytropic
equation of state with index $n=1$. We have shown the mass-radius diagram assuming
that strange stars are made of up of $(5-10)~\%$ of dark matter.
We conclude that if strange stars do exist, and if they accumulate dark matter, our findings limit in a certain way the radius and the mass of these compact objects.

After that we studied the radial oscillations of dark matter admixed strange stars.
Integrating numerically the equations for the perturbations we solved the corresponding boundary value problem to compute the first 11 frequency radial modes for three stars with the same mass and radius, but with different dark matter amounts. The large frequency separation were computed as well, and we showed them for all three models in the same plot for comparison reasons so that the impact of dark matter could be inferred.

In the second part we studied radial oscillations of Dark BEC stars made of ultralight repulsive scalar particles in the Fermi-Thomas approximation. Using the known background solution to the Lane-Emden equation for a Newtonian polytropic star with index $n=1$ we solved the Sturm-Liouville boundary value problem for the perturbation with the shooting method. We have computed the fundamental as well as several excited modes for two different star masses, and we have shown graphically i) several eigenfunctions corresponding to the first three and two highly excited oscillation modes, and ii) how the large frequency difference varies with the frequencies themselves. In addition, we have reformulated the boundary value problem equivalently by writing down a Schr{\"o}dinder-like equation, and we have shown the effective potential together with the first five values of $\omega^2$.

Finally, we have studied non-radial oscillations of bosonic configurations made of ultralight repulsive scalar particles in the Cowling approximation. For three different values of the angular degree $l=1,2,3$ we have computed the lowest frequencies, several associated eigenfunctions, and the effective potential in the equivalent description in terms of a Schr{\"o}dinder-like equation. The large frequency separations are shown as well. In all three cases, like in the radial oscillation case, for the higher excited modes the large separation tends to a constant determined entirely by the mass scale $\Lambda=\sqrt{m F}$, where $m$ is the mass of the scalar field and $F$ is a high mass scale that determines the self-interaction coupling constant in the scalar potential.



\section*{Acknowledgements}

I wish to thank the organizers of the Bled Workshop for their kind invitation to participate and present my work. I also thank the Funda\c c\~ao para a Ci\^encia e Tecnologia (FCT), Portugal, for the financial support to the Center for Astrophysics and Gravitation-CENTRA, Instituto Superior T\'ecnico,  Universidade de Lisboa, through the Grant No. UID/FIS/00099/2013.


\end{document}